\documentclass[preprint,nofootinbib]{revtex4}

\usepackage{amsmath}
\usepackage{graphicx}

\newcommand{\KS}{{K_SK_SK_S}}
\newcommand{\SK}{{\mathcal S}(K^0K^0\overline{K}{}^0)}
\newcommand{\xinu}{\xi_\nu}
\newcommand{\axi}{\left|\xi\right|}
\newcommand{\axinu}{\left|\xinu\right|}
\newcommand{\xib}{\overline{\xi}}
\newcommand{\axib}{\left|\xib\right|}
\newcommand{\axibh}{\left|\widehat{\xib}\right|}
\newcommand{\su}{SU(3)}
\newcommand{\ket}[1]{\left|#1\right\rangle}

\newcommand{\braket}[2]{\left\langle #1 | #2 \right\rangle}

\newcommand{\K}{K^0}
\newcommand{\Kb}{\overline{K}{}^0}
\newcommand{\Sym}[1]{\mathcal{S}(#1)}
\newcommand{\lamsc}{V^*_{cb}V_{cs}}
\newcommand{\lamsu}{V^*_{ub}V_{us}}
\newcommand{\lamdc}{V^*_{cb}V_{cd}}
\newcommand{\lamdu}{V^*_{ub}V_{ud}}
\newcommand{\pq}{\left(V^*_{cs}V_{cd}\right)/\left|V^*_{cs}V_{cd}\right|}
\renewcommand{\Re}{\mathcal{R}e}
\renewcommand{\Im}{\mathcal{I}m}

\newcommand{\beq}{\begin{equation}}
\newcommand{\eeq}{\end{equation}}
\newcommand{\beqa}{\begin{eqnarray}}
\newcommand{\eeqa}{\end{eqnarray}}
\newcommand{\no}{\nonumber}
\def\OMIT#1{{}}
\newcommand{\lsim}{\mathrel{\rlap{\lower4pt\hbox{\hskip1pt$\sim$}}
    \raise1pt\hbox{$<$}}}         
\newcommand{\gsim}{\mathrel{\rlap{\lower4pt\hbox{\hskip1pt$\sim$}}
    \raise1pt\hbox{$>$}}}         

\begin{document}

\preprint{{\vbox{\hbox{}\hbox{}\hbox{}
    \hbox{WIS/10/05-May-DPP}
\hbox{hep-ph/0505194}}}}

\vspace*{1.5cm}

\title{SU(3) Relations and the CP Asymmetry in $B\to K_SK_SK_S$}

\author{Guy Engelhard}\email{guy.engelhard@weizmann.ac.il}
\affiliation{Department of Particle Physics,
  Weizmann Institute of Science, Rehovot 76100, Israel}
\author{Yosef Nir}\email{yosef.nir@weizmann.ac.il}
\affiliation{Department of Particle Physics,
  Weizmann Institute of Science, Rehovot 76100, Israel}
\author{Guy Raz}\email{guy.raz@weizmann.ac.il}
\affiliation{Department of Particle Physics,
  Weizmann Institute of Science, Rehovot 76100, Israel}


\vspace{2cm}
\begin{abstract}
  The CP asymmetry in the $B\to K_S K_S K_S$ decay is being measured
  by the two B factories. A large deviation of the CP asymmetry
  $S_{K_SK_SK_S}$ from $-S_{\psi K_S}$ and/or of $C_{K_SK_SK_S}$ from
  zero would imply new physics in $b\to s$ transitions. We try to put
  upper bounds on the Standard Model size of these deviations, using
  SU(3) flavor relations and experimental data on the branching ratios
  of various decay modes that proceed via $b\to d$ transitions. We
  point out several subtleties that distinguish the case of three body
  final states from two body ones. We present several simple relations
  that can become useful once all relevant modes are measured
  accurately enough. 
\end{abstract}

\maketitle

\section{Introduction}
\label{sec:introduction}
The Belle \cite{Abe:2004sx} and Babar \cite{Aubert:2005dy} experiments
have recently presented their first results on the CP asymmetries in
$B\to\KS$ decays. The average of the two measurements is given by
\cite{HFAG} 
\begin{equation}
  \label{eq:1}
  \begin{aligned}
    S_{\KS}&=&-0.26\pm0.34,\\
    C_{\KS}&=&-0.41\pm0.22.
  \end{aligned}
\end{equation}
The $B\to\KS$ decay is a flavor changing neutral current process and,
consequently, does not proceed via tree level diagrams. Within the
Standard Model, the $b\to s$ penguin contributions are dominated by a
single weak phase, that is the phase of $V_{cb}^*V_{cs}$. The effects
of a second phase, that is the phase of $V_{ub}^*V_{us}$, are CKM
suppressed by ${\cal O}(\lambda^2)$. Neglecting the latter contributions,
and taking into account that the $\KS$ state is purely CP even
\cite{Gershon:2004tk}, the SM predictions are then as follows:
\beqa\label{eq:2}
S_\KS&\approx&-S_{\psi K_S}\no\\
C_\KS&\approx&0,
\eeqa
where, experimentally, $S_{\psi K_S}=0.726\pm0.037$ \cite{HFAG}.
These predictions are valid also in extensions of the SM where the
$B^0-\overline{B}^0$ mixing amplitude is possibly affected by new
phases, but the $b\to s$ decay amplitudes are not.

In order to understand whether violations of eq.~\eqref{eq:2} signal
new physics, it is necessary to estimate or, at least, put an upper
bound on the CKM-suppressed SM contributions. Such a calculation
involves, however, in addition to the CKM factors, hadronic physics.  
Currently, no first principle method for calculating hadronic matrix
elements has been proven to work to a high level of
precision. Furthermore, existing methods (for example,
\cite{Ali:1998eb,Beneke:2001ev,Beneke:2003zv,Bauer:2004tj}) 
have only been applied to two body final states, while our interest
here lies in the three body mode $B\to\KS$.

In this work we use the approximate $\su$ of the strong interactions
to constrain the relevant hadronic matrix elements
\cite{Grossman:2003qp}. While this method has the advantage of being
hadronic-model independent, it has the following two
weaknesses. First, $\su$ breaking effects can be of order 30\%, so that
our results cannot be trusted to better accuracy than that. Second,
since we have no information about the strong phases that are
involved, we make the most conservative assumption, whereby all
amplitudes interfere constructively. This leads to upper bounds on the
deviations from eq.~\eqref{eq:2} that are often much weaker than the
actual deviations expected in the Standard Model. Furthermore, the
quality of our upper bounds depends on the precision of current
measurements. Thus, one should not think of our bounds as estimates of
the deviation expected within the Standard Model. They are only
approximate (to ${\cal O}(0.3)$) and, in most cases, very conservative
upper bounds (with the advantage of being model independent).

This type of analysis has been previously applied to CP asymmetries in
decays into two body final states using the full $\su$ symmetry or
into three body final states using an $SU(2)$ subgroup
\cite{Grossman:1997gr,Grossman:2003qp,Gronau:2003ep,Chiang:2003pm,Gronau:2003kx,Gronau:2004hp}.
For the mode of interest to us, an $SU(2)$ analysis is not enough. We
thus study three body decays in the framework of the full $\su$ group.

The analysis of three body final states involves several subtleties and
technical complications. We have developed methods to overcome these
difficulties that are of more general applicability than just the
$\KS$ mode. As concerns our final results, we find that current
experimental data give no constraint on the CP asymmetry in $B \to
\KS$ using only $\su$. It is possible, however, that future
measurements of branching ratios of a few additional three body modes,
together with an improvement in the constraints on a few other, will
lead to useful constraints.  (When we add a rather mild dynamical
assumption to our $\su$ analysis, we do obtain a bound with present
data. The experimental range of the CP asymmetries is consistent with
this bound.)

The plan of this paper is as follows. In
section~\ref{sec:constr-cp-asymm} we introduce formalism and
notations that are specifically suitable for three body decays.
In section~\ref{sec:using-su-relations} we explain the principles of
how to obtain $\su$ relations that constrain the CP asymmetries in
three body decays. In section~\ref{sec:actual-su-relations} we focus
on the mode of interest, $B\to\KS$, and give a few concrete examples
of amplitude relations, as well as a Table that allows one to derive
all relevant relations. We conclude in section~\ref{sec:conclusions}.
Technical details are further discussed in two appendices. In
appendix~\ref{sec:cp-asymm-depend} we derive the relations between the
CP asymmetries and the parameters that we define in
section~\ref{sec:constr-cp-asymm}, and we justify the approximations
that we use. In appendix~\ref{sec:su-ampl-relat} we describe the
techniques that we developed to deal with the complicated $\su$
decomposition of the decay amplitudes. Appendix~\ref{sec:exp-data}
contains a list of relevant branching ratios. 

\section{Notations and Formalism}
\label{sec:constr-cp-asymm}
In this section we show how to modify and generalize the analysis
of ref.~\cite[]{Grossman:2003qp} so that it can be applied to three-body 
decays.

Unlike two body decays, the final state in three body decays is not
uniquely determined by the identity of the final mesons. Additional
quantum numbers (for example, the momenta) are needed to specify the
state. We use abstract vector notation, {\it e.g.} $\vec{A}_\KS$,
where the vector index runs over all possible values for the quantum
numbers, to describe the various states. The total decay rate is given by
\begin{equation}\label{eq:3}
  \Gamma(B^0 \to K_SK_SK_S)=\left\|\vec{A}_{K_SK_SK_S}\right\|^2\;.
\end{equation}
This equation defines the normalization of the decay amplitudes
$\vec{A}_f$. The norm in the right hand side of eq.~\eqref{eq:3}
represents a sum over all possible final states. If we choose to
describe the different final states using definite linear momenta, the
norm is actually calculated by an integral over all momentum
configurations. We stress that the norm is the same, no matter which 
basis we choose to span the final states with.

In order to derive $\su$ relations, we choose to span the final
states in a basis with definite linear momenta. Our convention is that
the order in which we write the three final mesons corresponds to
their momentum configuration:
\begin{equation}
  \left|M_iM_jM_k\right\rangle\equiv
\ket{M_i(p_1)M_j(p_2)M_k(p_3)}.\label{eq:4}
\end{equation}
We further define symmetrized states, $\ket{\mathcal{S}(f)}$, as follows:
\begin{equation}
  \begin{split}
  \ket{\mathcal{S}(M_1M_1M_1)} & \equiv \ket{M_1M_1M_1} \;, \\
  \ket{\mathcal{S}(M_1M_1M_2)} & \equiv \frac{1}{\sqrt{3}} 
  \left(\ket{M_1M_1M_2} + \ket{M_1M_2M_1} + \ket{M_2M_1M_1}\right) \;,
  \\ 
  \ket{\mathcal{S}(M_1M_2M_3)} & \equiv \frac{1}{\sqrt{6}} \left(
    \ket{M_1M_2M_3} + \ket{M_2M_3M_1} + \ket{M_3M_1M_2} \right. \\
    & \qquad +
    \left. \ket{M_3M_2M_1} + \ket{M_2M_1M_3} + \ket{M_1M_3M_2}
    \right)\;.
\end{split}\label{eq:5}
\end{equation}
In eq.~\eqref{eq:5}, $M_1$, $M_2$ and $M_3$ stand for different
mesons.  

Focussing on the mode of interest to us, namely $B^0$ decay into a
final $\ket{\KS}$ state, we note it can proceed via any of the three
$\Delta s=-1$ transitions whereby a $B^0$ meson decays into
$\ket{\K\K\Kb}$, $\ket{\K \Kb \K}$ or $\ket{\Kb \K\K}$. Owing to the
symmetry of the $\ket{\KS}$ state under exchange of any two of the
final mesons, it can only come from the totally symmetric combination
of the three states, $\ket{\mathcal{S}(\K \K \Kb)}$.
Neglecting CP violation in the neutral kaon mixing (the experimental
measurement of $\varepsilon_K$ guarantees that this approximation is
good to ${\cal O}(10^{-3})$), we have
\begin{equation}\label{eq:6}
  \braket{\KS}{\mathcal{S}(\K \K\Kb)}=\sqrt{\frac{3}{8}}
  \frac{V_{cs}^*V_{cd}}{|V_{cs}^*V_{cd}|}\;,
\end{equation}
for every set of values for the momenta $p_1,p_2,p_3$.
There are two additional combinations of $\ket{\K \K \Kb}$, $\ket{\K
  \Kb \K}$ and $\ket{\Kb \K \K}$ which are orthogonal to
$\ket{\mathcal{S}(\K \K \Kb)}$. However, since the projection of these
combinations on $\ket{\KS}$ is zero, we can write
\begin{equation}\label{eq:7}
  \vec{A}_\KS = \sqrt{3/8}\,\left[\pq\right]\,
  \vec{A}_{\SK}\;.
\end{equation}

Within the Standard Model, the violation of CP is encoded in the
complex phases of the CKM elements. It is therefore convenient, for
the purpose of discussing CP asymmetries, to have the CKM dependence
explicit. Following the discussion above, we thus
write the $B^0\to K_SK_SK_S$ decay amplitudes as follows: 
\begin{equation}\label{eq:8}
  \vec{A}_{K_SK_SK_S}= \left(\lamsc\, \vec{a}^c_{\SK} +
  \lamsu\, \vec{a}^u_{\SK}\right)\sqrt{3/8}\left[\pq\right]\;.
\end{equation}
Here, and for all other processes discussed below, the amplitudes for
the CP-conjugate processes, $\overline{B}{}^0\to \overline{f}$, have
the CKM factors complex-conjugated, while the $\vec{a}^{u,c}_f$
factors remain the same. 

Generalizing~\cite{Grossman:2003qp}, we introduce a parameter $\xi$: 
\begin{equation}\label{eq:9}
\xi\equiv \frac{\left|\lamsu\right|}{\left|\lamsc\right|}\,
  \frac{\vec{a}^c_{\SK} \cdot \vec{a}^u_{\SK}}{\|\vec{a}^c_{\SK}\|^2}\;,
\end{equation}
where the dot product of complex vectors is defined by
$\vec{X}\cdot \vec{Y} \equiv \sum_\nu X_\nu^*\,Y_\nu$.
Another useful parameter, $\axib$, is defined as follows:
\begin{equation}
  \label{eq:10}
  \axib \equiv \frac{\left|\lamsu\right|}{\left|\lamsc\right|}\,
  \frac{\|\vec{a}^u_{\SK}\|}{\|\vec{a}^c_{\SK}\|}\;.
\end{equation}
We have
\begin{equation}
  \label{eq:11}
  \frac{\axi}{\axib}=\frac{|\vec{a}^c_{\SK} \cdot
  \vec{a}^u_{\SK}|}{\|\vec{a}^c_{\SK}\| \cdot \|\vec{a}^u_{\SK}\|}\leq1 \;.
\end{equation}
The parameter $\axib$ is the one which can be constrained by $\su$
relations, and that would lead, through eq. (\ref{eq:11}), to a
constraint on $\axi$.

The case of two body decays \cite{Grossman:2003qp} constitutes a
specific example of our more general notation~(\ref{eq:9}), where the
vectors are simply one dimensional and, as can be seen from
eq. (\ref{eq:11}), $\axi=\axib$. The way in which $\xi$ of ref. 
\cite{Grossman:2003qp} is defined differs, however, by a weak phase
factor:
$\xi$(ref. \cite{Grossman:2003qp})$=e^{i\gamma}\xi$(eq. \ref{eq:9}).

Before concluding this section, we introduce one more
definition. Experiments often measure charge-averaged rates,
\beq\label{chaave}
\Gamma(B\to f)=\frac12\left[\Gamma(B^0\to
  f)+\Gamma(\overline{B}^0\to\overline{f})\right],
\eeq
where $\overline{f}$ is the CP-conjugate state of $f$. For CP
eigenstates, $\overline{f}=\pm f$. When a single weak phase dominates,
the CP-conjugate rates are equal, $\Gamma(B^0\to
f)=\Gamma(\overline{B}^0\to\overline{f})$, and there is no reason to
make a distinction between $\Gamma(B\to f)$ and $\Gamma(B^0\to f)$.

\section{Constraining the CP Asymmetries}
\label{sec:using-su-relations}
As we show in appendix~\ref{sec:cp-asymm-depend}, we can write, to
first order in $\Re(\xi)$ and $\Im(\xi)$,
\begin{align}\label{eq:12}
  -S_\KS - S_{\psi K_S} & = 2\, \cos 2\beta\, \sin
  \gamma\, \Re(\xi)\;, \\
  \label{eq:13}
  C_\KS & = - 2\, \sin\gamma\, \Im(\xi)\;.
\end{align}
The significance of the parameter $\xi$ is that it encodes all
hadronic physics that affects the deviation of $-S_\KS$ from
$\sin2\beta$ and of $C_\KS$ from zero. The other parameters, $\beta$
and $\gamma$, are weak phases that can be determined rather accurately
from other measurements. We learn that if we are able to put an upper
bound, $|\xi|\leq|\xi|^{\max}$, we will obtain an unambiguous test of the Standard
Model CP violation by asking whether the relation \cite{Grossman:2003qp}
\beq\label{eq:cs}
\left[(S_\KS +S_{\psi K_S})/\cos 2\beta\right]^2+
C_\KS^2\leq4\sin^2\gamma(|\xi|^{\rm max})^2
\eeq
is fulfilled. As mentioned above, the parameter that appears in
the $\su$ relations is actually $\axib$. In this and the next section,
we assume that $\su$ is exact, and use it to constrain $|\xi|$.

In order to constrain $\axib$ we consider $\Delta s=0$ decay
amplitudes and write, using our vector notation,
\begin{equation}\label{eq:14}
    \vec{A}_{f}=V^*_{cb}V_{cd}\, \vec{b}^c_{f}+V^*_{ub}V_{ud}\, \vec{b}^u_{f}\;.
\end{equation}
$\su$ relations lead to amplitude relations of the form
\begin{equation}
  \label{eq:15}
  \vec{a}^q_{\SK} = \sum\limits_f X'_f\,
  \vec{b}^q_f\; \ \ \ (q=u\ {\rm or}\ c).
\end{equation}
Most generally, in the sum over $f$, states which are permutations of
each other are treated as different states (for example,
$K^+K^-\pi^0$ and $K^+\pi^0K^-$ are different states). However, since
the state $\ket{\SK}$ is completely symmetric, the strongest
constraint is obtained from summing only over completely symmetric
terms. Making this choice for eq.~\eqref{eq:15} allows us to
rewrite it as follows: 
\begin{equation}\label{eq:16}
  \vec{a}^q_{\SK} = \sum\limits_f X_f\, \vec{b}^q_{\mathcal{S}(f)}\;.
\end{equation}
The $X_f$'s of eq.~\eqref{eq:16} are related to the $X'_f$'s of
eq.~\eqref{eq:15} by symmetry factors. Taking the norm
of eq.~\eqref{eq:15} needs to be done with care: the sum can
involve states with different symmetry properties, and the
corresponding norms have different meanings. On the other hand, there
is no ambiguity in taking the norm of eq.~\eqref{eq:16}. Consequently, we
can write 
\begin{equation}\label{eq:17}
  \left\|\vec{a}^q_{\SK}\right\| \leq \sum\limits_f
  \left|X_f\right|\,\left\|\vec{b}^q_{\mathcal{S}(f)}\right\|\;.
\end{equation}

We provisionally assume, for simplicity, that $\Delta s=\pm1$ decays are
dominated by the $\vec{a}^c$ terms (this assumption is justified if
$\axib$ is small, see eq.~\eqref{eq:10}), while $\Delta
s=0$ decays are dominated by the $\vec{b}^u$ terms. (Below we obtain our
constraints without making these assumptions, in a fashion similar
to~\cite{Grossman:2003qp}.) Then the amplitudes are related to the
decay rates by
\begin{align}
  \label{eq:18}
  \left|\lamsc\right| \left\|\vec{a}^c_{\SK}\right\| &
  \approx \sqrt{(8/3)\Gamma({B}\to K_SK_SK_S)}\;, \\
\label{eq:19}
  \left|\lamdu\right|\left\|\vec{b}^u_{\mathcal{S}(f)}\right\| &
  \lesssim \sqrt{\Gamma({B}\to f)}\;.
\end{align}
The inequality in eq.~\eqref{eq:19} comes from the fact that we
consider the symmetrized state, rather then a generic state, for which
$\left|\lamdu\right|\left\|\vec{b}^u_{f}\right\|\approx
\sqrt{\Gamma({B}\to f)}$. Combining eqs.~\eqref{eq:17},
\eqref{eq:18} and~\eqref{eq:19}, we get  
\begin{equation}
  \label{eq:20}
  \axib \lesssim  \left|\frac{V_{us}}{V_{ud}}\right|
  \sum\limits_f  \left|X_f\right| \sqrt{\frac{\Gamma({B}\to
      f)}{(8/3)\, \Gamma({B}\to K_SK_SK_S)}}\;.
\end{equation}

We now proceed without making the assumptions of $\vec{a}^c_f$- and
$\vec{b}^u_f$-dominance which led to eqs.~\eqref{eq:18}
and~\eqref{eq:19} \cite[]{Grossman:2003qp}. Instead of
$\axib$, we constrain a new parameter, $\axibh$, defined by
\begin{equation}\label{eq:21}
  \axibh^2 \equiv \left|\frac{V_{us}}{V_{ud}}\right|^2
  {\tfrac{\left\| \lamdc\,  \vec{a}^c_{\SK}
      +\lamdu\,\vec{a}^u_{\SK} \right\|^2+\left\| V_{cb}V^*_{cd}\,  \vec{a}^c_{\SK}
      +V_{ub}V^*_{ud}\,\vec{a}^u_{\SK} \right\|^2}
  {\left\| \lamsc\,  \vec{a}^c_{\SK}
      +\lamsu\,\vec{a}^u_{\SK} \right\|^2+\left\| V_{cb}V^*_{cs}\,  \vec{a}^c_{\SK}
      +V_{ub}V^*_{us}\,\vec{a}^u_{\SK} \right\|^2}}\;.
\end{equation}
The numerator and denominator of $\axibh^2$ are related to
charge-averaged rates:
\begin{align}\label{eq:22}
  \begin{split}
    \left\| \lamdc\, \vec{a}^c_{\SK}+\lamdu\,\vec{a}^u_{\SK}
    \right\|^2 & +\left\| V_{cb}V^*_{cd}\, \vec{a}^c_{\SK}
      +V_{ub}V^*_{ud}\,\vec{a}^u_{\SK} \right\|^2 \\
    & \leq 2\left(\sum\limits_f\left|X_f\right| \sqrt{\Gamma({B}\to f)}\right)^2\;,
  \end{split} \\
  \label{eq:23}
  \begin{split}
    \left\| \lamsc\, \vec{a}^c_{\SK}+\lamsu\,\vec{a}^u_{\SK}
    \right\|^2 & +\left\| V_{cb}V^*_{cs}\, \vec{a}^c_{\SK}
      +V_{ub}V^*_{us}\,\vec{a}^u_{\SK} \right\|^2 \\
    & = (16/3)\Gamma({B}\to K_SK_SK_S)\;.
  \end{split}
\end{align}
Using the measured charge-averaged rates, a constraint on $\axibh^2$ is
obtained without any further assumptions.

\begin{figure}
  \begin{minipage}{0.5\linewidth}
    \begin{flushright}
      \raisebox{0.8\linewidth}{$\axib$} \includegraphics[height=0.8\linewidth]{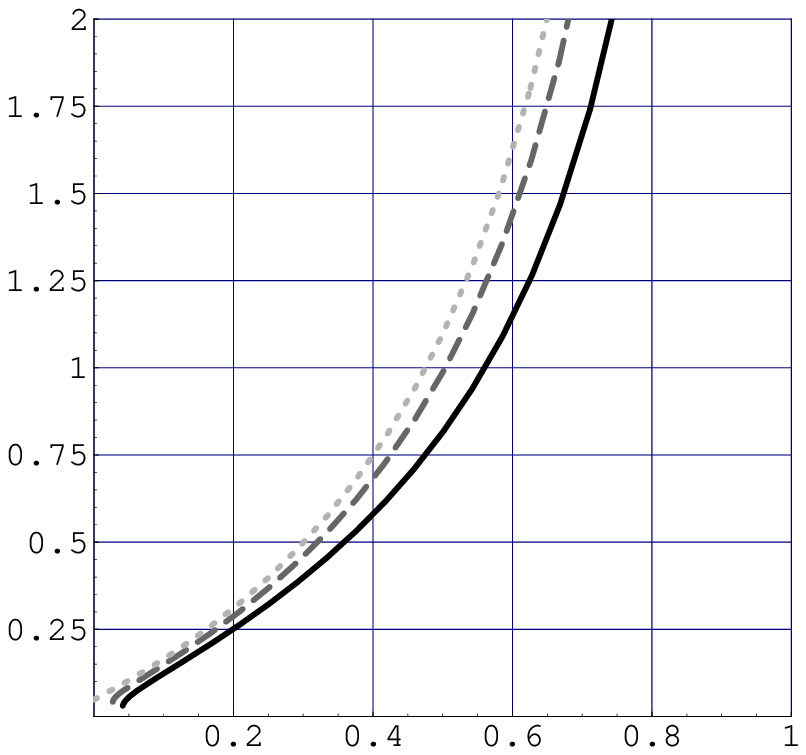}
      \phantom{$\axib$} \\
      $\axibh$
    \end{flushright}
    \end{minipage}
    \caption{The upper bound that can be placed on $\axib$ as
      a function of the upper bound on $\axibh$, according to
      eq.~(\ref{eq:24}). The three curves correspond to different ways
      of treating the weak phase $\gamma$: $\gamma=59.0^\circ$, the
      experimental central value (solid black); $\gamma \in
      [35.9^\circ,80.1^\circ ]$, the $3\sigma$
      range~\cite{Charles:2004jd} (dark-gray dashed); $\gamma$
      unconstrained (light-gray dotted).} 
  \label{fig:xihat}
\end{figure}

The $\axibh$ and $\axib$ parameters are related as follows:
\begin{equation}\label{eq:24}
\axibh^2 = \frac{ \left|\frac{V_{us}V_{cd}}{V_{cs}V_{ud}} \right|^2
+\axib^2 + 2\cos\gamma\,\Re\left(\frac{V_{us}V_{cd}}{V_{cs}V_{ud}}\,\xi\right)}{1 +
\axib^2 + 2 \cos\gamma\,\Re(\xi)}\;.
\end{equation}
The relation~\eqref{eq:24} is a generalization of the relation
in~\cite[eq.
14]{Grossman:2003qp}.\footnote{Ref.~\cite[]{Grossman:2003qp} uses the
  rates $\Gamma(B^0\to f)$ 
  to define $\left|\hat{\xi}\right|$, while we use the charge-averaged
  rates $\Gamma(B \to f)$. If we leave $\gamma$ unconstrained, the
  resulting upper bound on $\axib$ is the same, but for $\gamma\neq0$,
  our expression gives a stronger bound, as can be seen in
  Fig.~\ref{fig:xihat}.} It has the important
property that for $\lambda^2 \lesssim \axibh\leq 1$ we get a
constraint on $\axib$, for any $\xi$ (of course, within the allowed
range, $|\xi|\leq\axib$, see eq.~\eqref{eq:11}). Since we do not know
the value of $\xi$, we should consider the weakest constraint, which
corresponds to $\Re(\xi)=\axib$ (the $(V_{us}V_{cd})/(V_{cs}V_{ud})$
term is experimentally known to be real to a good approximation). We
show in Fig.~\ref{fig:xihat} the relation between the upper bound on
$\axibh$ and the resulting upper bound on $\axib$, for three ranges of
the weak phase $\gamma$. The weakest bound, which corresponds to
$\Re(\xi)=\axib$ and $\gamma=0$, is the curve
$\axibh=(\axib-\lambda^2)/(1+\axib)$. Note that the translation from
$\axibh$ to $\axib$ is non-linear, a point which was not stressed
in~\cite[]{Grossman:2003qp}, although it is true there as well.

\section{$\su$ relations for $\KS$}
\label{sec:actual-su-relations}
The simplest way to find $\su$ relations is to express the decay
amplitudes using invariant $\su$ reduced matrix elements.  While the
number of $\su$ independent reduced matrix elements in three body
decays is quite large, a significant simplification is obtained by the
fact that we only consider completely symmetric final
states. In particular, for generic $f$'s we have $40$ independent
reduced matrix elements, while for ${\mathcal S}(f)$'s there are only $7$. 

By scanning over all possible contractions of the relevant $\su$
tensors, we are able to obtain all matrix element relations in a
systematic way, avoiding the need to discuss $\su$ properties of
tensor products. We give more details on the calculation in
appendix~\ref{sec:su-ampl-relat}.

The main results of our work are summarized in
table~\ref{tab:matrix-elements} where we list the dependence of the
symmetrized three body decay amplitudes on $\su$ reduced matrix
elements. We give here only $B^0$ and $B^+$ decays, but it is
straightforward to add $B_s$ decays in a similar way. We stress that,
since this table includes only totally symmetric states, it is only
applicable to constrain the totally symmetric final states such as
$\KS$. 

\begin{table}[tb]
  \centering
  {\scriptsize
    \begin{tabular}{|c|c|c|c|c|c|c|c|}
      \hline
      $\mathcal{S}(f)$ & $A_1$ & $A_2$ & $A_3$ & $A_4$ & $A_5$ & $A_6$ & $A_7$\\
      \hline
      \hline
      $\mathcal{S}(K^0\bar K^0 K^0)$&$1$&$0$&$0$&$0$&$0$&$0$&$0$\\
      \hline
      \hline
      $\mathcal{S}(K^+K^-\pi^0)$&$0$&$1$&$0$&$0$&$0$&$0$&$0$\\
      $\mathcal{S}(K^0\bar K^0\pi^0)$&$0$&$0$&$1$&$0$&$0$&$0$&$0$\\
      $\mathcal{S}(\pi^+\pi^-\pi^0)$&$0$&$0$&$0$&$1$&$0$&$0$&$0$\\
      $\mathcal{S}(K^+\pi^-\bar K^0)$&$0$&$0$&$0$&$0$&$1$&$0$&$0$\\
      $\mathcal{S}(K^-\pi^+ K^0)$&$0$&$0$&$0$&$0$&$0$&$1$&$0$\\
      $\mathcal{S}(K^+K^-\pi^+)$&$0$&$0$&$0$&$0$&$0$&$0$&$1$\\
      $\mathcal{S}(K^+ \bar K^0 \pi^0)$&$0$&$-1$&$0$&$1$&$\frac{1}{\sqrt{2}}$&$0$&$0$\\
      $\mathcal{S}(K^0\bar K^0\pi^+)$&$0$&$0$&$-{\sqrt{2}}$&${\sqrt{2}}$&$0$&$-1$&$1$\\
      $\mathcal{S}(K^0\bar
      K^0\eta_8)$&$\frac{1}{\sqrt{3}}$&$0$&$\frac{1}{\sqrt{3}}$&$0$&$0$&$0$&$0$\\
      $\mathcal{S}(K^+K^-\eta_8)$&$\frac{1}{\sqrt{3}}$&$\frac{1}{\sqrt{3}}$&$
      \frac{2}{\sqrt{3}}$&$-\frac{2}{\sqrt{3}}$&$0$&$0$&$0$\\
      $\mathcal{S}(\pi^+\pi^-\eta_8)$&$\frac{1}{\sqrt{3}}$&$0$&$\frac{2}{\sqrt{3}}$&$
      -\frac{1}{\sqrt{3}}$&$\sqrt{\frac{2}{3}}$&$\sqrt{\frac{2}{3}}$&$0$\\
      $\mathcal{S}(K^+\bar K^0\eta_8)$&$0$&$-\frac{1}{\sqrt{3}}$&$0$&$
      \frac{1}{\sqrt{3}}$&$\frac{1}{\sqrt{6}}$&$0$&$0$\\
      $\mathcal{S}(\pi^+\pi^0\eta_8)$&$0$&$-\sqrt{\frac{2}{3}}$&$\sqrt{\frac{2}{3}}$&$0$&$
      \frac{1}{\sqrt{3}}$&$\frac{1}{\sqrt{3}}$&$0$\\
      \hline
      $\mathcal{S}(\pi^+\pi^+\pi^-)$&$0$&$0$&$0$&$0$&$0$&$0$&$\sqrt{2}$\\
      $\mathcal{S}(\pi^+\pi^0\pi^0)$&$0$&$0$&$0$&$0$&$0$&$0$&$\frac{1}{\sqrt{2}}$\\
      $\mathcal{S}(\pi^0\pi^0\eta_8)$&$\frac{1}{\sqrt{6}}$&$\sqrt{\frac{2}{3}}$&$0$&$
      -\frac{1}{\sqrt{6}}$&$0$&$0$&$0$\\
      $\mathcal{S}(\pi^0\eta_8\eta_8)$&$0$&$\frac{2\sqrt{2}}{3}$&$\frac{2\sqrt{2}}{3}$&$
      -\frac{5}{3\sqrt{2}}$&$0$&$0$&$0$\\
      $\mathcal{S}(\pi^+\eta_8\eta_8)$&$0$&$\frac{-2}{3}$&$\frac{-2}{3}$&$
      \frac{4}{3}$&$\frac{\sqrt{2}}{3}$&$
      -\frac{\sqrt{2}}{3}$&$\frac{1}{\sqrt{2}}$\\
      \hline
      $\mathcal{S}(\pi^0\pi^0\pi^0)$&$0$&$0$&$0$&$\sqrt{\frac{3}{2}}$&$0$&$0$&$0$\\
      $\mathcal{S}(\eta_8\eta_8\eta_8)$&$\frac{1}{\sqrt{2}}$&$\frac{\sqrt{2}}{3}$&$
      \frac{2\sqrt{2}}{3}$&$-\frac{1}{\sqrt{2}}$&$0$&$0$&$0$\\
      \hline
    \end{tabular}
  }
  \caption{$SU(3)$ decomposition of $A_{\mathcal{S}(f)}$. The
    different blocks refer to different degrees of symmetrization
    needed for each state.}
  \label{tab:matrix-elements}
\end{table}

A simple examination of table~\ref{tab:matrix-elements} reveals that
all the relations of the form of eq.~\eqref{eq:16} involve at least
one $\Delta s=0$ decay into a final state with an $\eta_8$ meson. In
the exact $\su$ limit, this corresponds to a state with a final $\eta$
meson. We would like to emphasize two points in this regard:
\begin{enumerate}
  \item The use of $\su$ relations involving amplitudes with final
    $\eta_8$ and/or $\eta_1$ mesons was recently criticized in
    ref.~\cite{Beneke:2005pu}, on the basis that $\su$ breaking
    effects in this system are large. The phenomenological value of
    the octet-singlet mixing angle is $\sin\theta\approx0.27$
    \cite{Feldmann:2002kz}. $\su$ breaking in the decay constants is
    parametrized by $2(f_s-f_q)/(f_s+f_q)\approx0.22$
    \cite{Feldmann:2002kz}. The breaking effects are thus consistent
    with our estimated accuracy of ${\cal O}(0.3)$, and are not ${\cal
      O}(1)$, as suggested in ref.~\cite{Beneke:2005pu}.
  \item If the relevant decays into final states involving
    $\eta^\prime$ are found to be enhanced compared to the
    corresponding states involving $\eta$, then the effects of the
    octet-singlet mixing on our results may be
    significant. These effects can, however, be taken into account by
    using both $\eta$ and $\eta^\prime$ data, in  a way similar to
    ref.~\cite{Grossman:2003qp}. 
\end{enumerate}

We now present several interesting specific relations. Note that,
since we do not know the values of the strong phases, in deriving our
bounds we must add the various $\Delta s=0$ amplitudes constructively,
see eq.~(\ref{eq:17}). This conservative procedure may weaken the
bound considerably. Therefore, relations involving a smaller number of
$\Delta s=0$ amplitudes are more likely to give strong bounds.

\subsection{A Single $\Delta s=0$ Amplitude}
There is no amplitude relation of the form of eq.~\eqref{eq:16} that
involves only a single $\Delta s=0$ decay amplitude of $B^0$ or
$B^+$. Such relations would have had the potential to lead to a tight
constraint. Note that, in general, we would still get an upper bound
on, rather than an estimate of, $\axibh$. The reason is that we consider
the symmetrized final state while experiments measure non-symmetrized
final states. However, there exists such a relation involving $B_s$
decays: 
\begin{equation}
  \label{eq:25}
  \vec{a}^q_{B_d \to \SK} = \vec{b}^q_{B_s \to \Sym{\Kb\Kb\K}}\;.
\end{equation}
This relation is, in fact, due to the U-spin subgroup of $\su$ and it
holds for the non-symmetrized states as well. Since in $B_s$ decays
the $\ket{\KS}$ state can only come from $\ket{\Sym{\Kb\Kb\K}}$, in a
way similar to eq.~(\ref{eq:7}), eq.~(\ref{eq:25}) implies relations
between the $B_d \to \KS$ and $B_s \to \KS$ decay amplitudes,
$V_{cs}V^*_{cd}\, \vec{a}^q_{B_d \to \KS} =
V_{cs}^*V_{cd}\,\vec{b}^q_{B_s \to \KS}$, leading to
\beq\label{eq:26}
\axibh=\left|\frac{V_{us}}{V_{ud}}\right|
\sqrt{\frac{\Gamma(B_s\to\KS)}{\Gamma(B_d\to\KS)}}.
\eeq

\subsection{Two $\Delta s=0$ Amplitudes}
We find a single amplitude relation involving only two $\Delta s=0$
amplitudes: 
\begin{equation}\label{eq:27}
  \vec{a}^q_{\Sym{\K\K\Kb}} = \sqrt{3}\,\vec{b}^q_{\Sym{\K\Kb\eta_8}}
  - \vec{b}^q_{\Sym{\K\Kb\pi^0}}\;.
\end{equation}
The fact that we are interested only in symmetrized states is helpful
here in yet another way. Let us write
\begin{equation}
  \label{eq:28}
  \braket{\Sym{\K \Kb X}}{\Sym{K_SK_S X}} =   - \braket{\Sym{\K \Kb
      X}}{\Sym{K_LK_L X}} = \frac{1}{\sqrt{2}}\;.
\end{equation}
($X$ here can be any meson except $\K$, $\Kb$, $K_S$ or $K_L$.) Since
in $B$ decays the $\ket{\Sym{K_SK_S X}}$ and $\ket{\Sym{K_LK_L X}}$
states can only come from an $\ket{\Sym{\K \Kb X}}$ state, we can write,
similarly to eq.~(\ref{eq:7}),
\begin{equation}\label{eq:29}
  \vec{A}_{\Sym{K_SK_SX}} = - \vec{A}_{\Sym{K_LK_LX}} =
  \frac{1}{\sqrt{2}}\,\vec{A}_{\Sym{\K \Kb X}}\;. 
\end{equation}
Consequently, the relation~(\ref{eq:27}) leads to the following
relation, which is more practical from the experimental point of view:
\begin{equation}\label{eq:30}
  \vec{a}^q_{\Sym{\K\K\Kb}} = \sqrt{6}\,\vec{b}^q_{\Sym{K_SK_S\eta_8}}
  - \sqrt{2}\,\vec{b}^q_{\Sym{K_SK_S\pi^0}}.
\end{equation}
($K_SK_S$ can be replaced by $K_LK_L$.) There is yet no measurement of
the modes in eq.~(\ref{eq:30}).

\subsection{Three $\Delta s=0$ Amplitudes}
We find several amplitude relations which involve three $\Delta s=0$
amplitudes, for example, 
\begin{align}\label{eq:31}
\vec{a}^q_{\Sym{\K\K\Kb}} = \sqrt{6}\,\vec{b}^q_{\Sym{\pi^0\pi^0\eta_8}} -
    2\,\vec{b}^q_{\Sym{K^+K^-\pi^0}} + \vec{b}^q_{\Sym{\pi^+\pi^-\pi^0}}\;.
\end{align}
At present, the branching ratio ${\cal B}({\pi^0\pi^0\eta})$ is not
yet constrained, while ${\cal B}({K^+K^-\pi^0})$ and ${\cal B}
({\pi^+\pi^-\pi^0})$ have rather weak upper bounds. For the relation
\eqref{eq:31} to become useful, the branching ratio of the first mode
must be constrained, and the bounds on the latter two must be improved.

\subsection{Measured $\Delta s=0$ Amplitudes}
There are relations which involve only modes which have been
measured. Given the experimental data in appendix~\ref{sec:exp-data},
the strongest bound is obtained by using the following relation:
\begin{equation}\label{eq:32}
  \begin{split}
    \vec{a}^q_{\SK} & = -\sqrt{2}\, \vec{b}^q_{\Sym{K^+ \pi^- \Kb}} +
    \sqrt{3}\,\vec{b}^q_{\Sym{\pi^+\pi^-\eta_8}} \\
    & \quad - \vec{b}^q_{\Sym{\pi^+\pi^-\pi^0}} - \sqrt{2}\,
    \vec{b}^q_{\Sym{K^+K^-\pi^+}} + \sqrt{2}\, 
    \vec{b}^q_{\Sym{\K \Kb \pi^+}}\;.
  \end{split}
\end{equation}
Using the definition~(\ref{eq:21}), we get
\begin{equation}\label{eq:33}
  \begin{split}
    \axibh \leq 0.22\,\sqrt{\frac{3}{8}} & \left(
      \sqrt{\frac{2\mathcal{B}(K^+\Kb\pi^-)}{\mathcal{B}(\KS)}} +
      \sqrt{\frac{3\mathcal{B}(\pi^+\pi^-\eta)}{\mathcal{B}(\KS)}}
      + \sqrt{\frac{\mathcal{B}(\pi^+\pi^-\pi^0)}{\mathcal{B}(\KS)}} 
    \right. \\
    & \quad \left.+ 
      \sqrt{\frac{2\mathcal{B}(K^+K^-\pi^+)}{\mathcal{B}(\KS)}}
      +  \sqrt{\frac{4\mathcal{B}(K_SK_S\pi^+)}{\mathcal{B}(\KS)}}
    \right) \leq 1.28\;.
  \end{split}
  \end{equation}
As explained above, we take here the $\su$ limit in replacing
$\eta_8$ with $\eta$. We also use $\mathcal{B}(\Sym{\K \Kb \pi^+})= 2\,
\mathcal{B}(\Sym{K_SK_S \pi^+})$.

We see that the strongest bound we can currently put on
$\axibh$ is too weak to bound $\axib$ and the CP asymmetries. It
is possible, however, that an improvement in experimental data, as
well as measurements of additional modes, will eventually lead to 
a significant bound.

\subsection{Dynamical Assumptions}
One can use simplifying dynamical assumptions and neglect the effect
of small contributions from exchange, annihilation, and penguin
annihilation diagrams~\cite{Dighe:1995bm}. Practically, this
means that all reduced matrix elements in which the spectator (the $B$
triplet) is contracted with a Hamiltonian operator are put to
zero. More details are given in Appendix~\ref{sec:su-ampl-relat}.
 
Such a simplification does lead to new relations. Most notably, there
is now an amplitude relation involving a single $\Delta s=0$ mode:
\begin{equation}\label{eq:34}
  \vec{a}^q_{\SK} = \sqrt{2}\, \vec{b}^q_{\Sym{\K\Kb\pi^+}}\;.
\end{equation}
This relation leads to the following upper bound: 
\begin{equation}\label{eq:35}
  \axibh \leq 0.22 \sqrt{\frac{3}{2}}
  \,\sqrt{\frac{\mathcal{B}(K_SK_S\pi^+)}{\mathcal{B}(\KS)}} \leq 0.20\;
  \ \ \Longrightarrow\ \ \  
  \axib \leq 0.31\;.
\end{equation}
The observed CP asymmetries are well within this bound.
We conclude that to uncover a signal of new physics with our 
methods will require improved experimental data.

\section{Conclusions}
\label{sec:conclusions}
We use the approximate $\su$ flavour symmetry to constrain the SM
pollution and the CP asymmetries in three body decays. This is an
extension of previous works that 
considered two body final states. One important difference is that 
two body final states are entirely defined by the identity of the final
mesons, while in three body decays additional quantum numbers (such as
momenta or angular momenta) are required to characterize the final
state. In the absence of an experimental spatial analysis, the
measured quantities are always a sum over all possible final
states. On the other hand, the $\su$ relations hold for each final
state separately. The application of $\su$ relations to three body
final states should therefore be done with care.

The case of $B \to \KS$ decay is special since the final state is
symmetric under the exchange of any two mesons. We showed
how this leads to a significant simplification in the $\su$ analysis,
allowing us to consider only final states with the same
symmetry and considerably reducing the number of independent $\su$
reduced matrix elements.

Still, decomposing the decay amplitudes for three body final states
into reduced matrix elements with well defined $\su$ transformation
properties is a difficult task in terms of group theoretical
calculation. Since, however, we are eventually interested only in
relations between physical decay amplitudes, we were able to scan
systematically over all possible reduced matrix elements and find an
independent subset of them. Using this method, our reduced matrix
elements bear no clean $\su$ interpretation, but their guaranteed
independence is all that matters for the task of finding amplitude
relations. The same method can be used to simplify other $\su$
calculations where the goal is obtaining physical amplitude relations.

Whether a numerical upper bound is achieved (and whether this
bound is strong enough to be in conflict with a measured CP
asymmetry) depends on the available experimental data. Currently, no
such bound can be obtained with no additional assumptions, and
the bound which is obtained when additional dynamical assumptions
are used is not strong enough to be in conflict with the measured CP
asymmetry. However, our work shows which new measurements have the
potential to lead to a constraint.

The hope is that, given more and better experimental data, three body
decays and $\su$ relations will provide us with an additional
unambiguous test of the SM mechanism of CP violation.

\acknowledgments
We are grateful to Zoltan Ligeti for his substantial contributions to
this work. We thank Yuval Grossman and Helen Quinn for their comments
on the manuscript.
This work was supported by a grant from the G.I.F., the
German--Israeli Foundation for Scientific Research and Development,
by the Israel Science Foundation
founded by the Israel Academy of Sciences and Humanities, by EEC RTN
contract HPRN-CT-00292-2002, by the Minerva Foundation (M\"unchen),
and by the United States-Israel Binational Science Foundation (BSF),
Jerusalem, Israel.

\appendix
\section{The $\xi$ dependence of $S_\KS$ and $C_\KS$}
\label{sec:cp-asymm-depend}
In order to derive the dependence of CP asymmetries in $B\to\KS$ on
$\xi$, we work in the basis of states with definite angular momentum
between two of the three final $K_S$ mesons. While it is difficult
to write $\su$ relations in this basis, the advantage of using it is
that any basis-state is manifestly CP even~\cite{Gershon:2004tk}. (In
other bases, final states which are related by CP would have related
decay amplitudes such that only the combinations which correspond to
CP even final states have a non-zero amplitude. The discussion,
however, is much simpler in the basis we choose.)

We denote the various components of the vector $\vec{A}_{K_SK_SK_S}$
in this basis by $A_{\nu}$ where $\nu$ is some collective index
which runs over all possible final states. We define for every
final state the parameter
\begin{equation}\label{eq:36}
\lambda_{\nu} \equiv e^{-i\phi_B}\, \frac{\bar{A}_{\nu}}{A_{\nu}}\;,
\end{equation}
where $\phi_B$ is the phase of the $B^0-\overline{B}{}^0$ mixing
amplitude (see the review on CP violation in \cite{Eidelman:2004wy}
and, in particular, eq. (58)). 

Currently, the experimental time dependent CP asymmetry is measured
with no distinction between various final $\KS$ states. The
expression for the measured CP asymmetry involves therefore a sum
over all final $\KS$ states:
\begin{equation}\label{eq:37}
  {\cal A}_{\KS}(t) =\frac{\sum\limits_\nu\Gamma(\overline{B}{}^0(t)\to \nu) -
    \sum\limits_\nu\Gamma(B^0(t)\to \nu)}
  {\sum\limits_\nu\Gamma(\overline{B}{}^0(t)\to \nu) +
    \sum\limits_\nu\Gamma(B^0(t)\to \nu)}\;.
\end{equation}
Writing
\begin{equation}\label{eq:38}
  {\cal A}_{\KS}(t) = S_{\KS} \sin(\Delta m\:t) - C_{\KS} \cos(\Delta m\:t)\;,
\end{equation}
we use the definition~\eqref{eq:36} to get
\begin{align}
  \label{eq:39}
  S_{K_SK_SK_S} &= \frac{\sum\limits_\nu \left|A_{\nu}\right|^2 2\,
    \text{Im} \lambda_{\nu}}{\sum\limits_\nu \left|A_{\nu}\right|^2
    \left(1+\left|\lambda_{\nu}\right|^2\right)}\;, \\
  \label{eq:40}
  C_{K_SK_SK_S} &= \frac{\sum\limits_\nu \left|A_{\nu}\right|^2
    \left(1-\left|\lambda_{\nu}\right|^2\right)}{\sum\limits_\nu
    \left|A_{\nu}\right|^2
    \left(1+\left|\lambda_{\nu}\right|^2\right)}\;.
\end{align}

We can make the CKM dependence of each amplitude explicit:
\begin{equation}\label{eq:41}
    A_{\nu}=\left(\lamsc a^c_{\nu}+\lamsu a^u_{\nu}\right)\left[\pq\right]
  = A^c_{\nu}\left(1+\xinu\,e^{i\,\gamma}\right)\;,
\end{equation}
where $\gamma$ is the weak phase between $\lamsc$ and $\lamsu$, and
$\xinu$ therefore contains only strong phases and is defined by
\begin{equation}
  \label{eq:42}
  \xinu \equiv
  \left|\frac{\lamsu}{\lamsc}\right|\,
  \frac{a^u_{\nu}}{a^c_{\nu}}\;. 
\end{equation}

Assuming that $\axinu$ is small for every $\nu$ (we justify this
assumption below), we expand~\eqref{eq:39} and~\eqref{eq:40} to
first order in $\axinu$:
\begin{align}
  \label{eq:43}
    S_{K_SK_SK_S} &= \sin2\beta\frac{\sum\limits_\nu \left|A^c_{\nu}\right|^2
    \eta_{\nu}}{\sum\limits_\nu \left|A^c_{\nu}\right|^2}
  + 2\, \cos 2\beta\, \sin \gamma\,
  \frac{\sum\limits_\nu \left|A^c_{\nu}\right|^2
    \eta_{\nu}\, \Re\left(\xinu
    \right)}{\sum\limits_\nu \left|A^c_{\nu}\right|^2}\;, \\ 
  \label{eq:44}
  C_\KS & = -2\, \sin \gamma\, \frac{\sum\limits_\nu \left|A^c_{\nu}\right|^2
    \Im\left(\xinu\right)}{\sum\limits_\nu \left|A^c_{\nu}\right|^2}\;.
\end{align}
At this point, the fact that all final states are CP even plays an
important role as it dictates that $\eta_\nu=1$ for all $\nu$.
Switching now to vector notation we have
\begin{align}
  \label{eq:45}
  \sum\limits_\nu\left|A^c_\nu\right|^2 & = \left|\lamsc\right|^2 \,
  \left\|\vec{a}^c\right\|^2\;, \\
  \label{eq:46}
  \sum\limits_\nu\left|A^c_\nu\right|^2\Re\left(\xinu\right) & =
  \left| \lamsc\,\lamsu\right| \mathcal{R}e
  (\vec{a}^c\cdot \vec{a}^u)\;, \\
  \label{eq:47}
  \sum\limits_\nu\left|A^c_\nu\right|^2\Im\left(\xinu\right) & =
  \left| \lamsc\,\lamsu\right| \mathcal{I}m
  (\vec{a}^c\cdot \vec{a}^u)\;.
\end{align}
Using the definition~\eqref{eq:9} we therefore get eqs.~\eqref{eq:12}
and~\eqref{eq:13} to first order in $\xi$. 

We still need to justify the expansions~\eqref{eq:43}
and~\eqref{eq:44}. In making it, we assumed that for every $\nu$ we
have $\axinu<1$. Unlike in two body decay, the mere smallness
of $\axib$ is not enough to validate this assumption for every
$\nu$. Nevertheless, we show next that the smallness of $\axib$ does
guarantee that the branching ratio of final states in which $\axinu
\geq 1$ is constrained by $\mathcal{O}(\axib^2)$. The terms that are
omitted in writing eqs.~\eqref{eq:43} and~\eqref{eq:44} are therefore
of $\mathcal{O}(\axib^2)$. 

Our starting point is the definition of $\axib$, eq.~\eqref{eq:10},
leading to
\begin{equation}
  \label{eq:48}
  \axib^2 =
  \frac{\left|\lamsu\right|^2}{\left|\lamsc\right|^2}
  \frac{\left\|\vec{a}^u\right\|^2}{\left\|\vec{a}^c\right\|^2}
  = \frac{\left|\lamsu\right|^2\left\|\vec{a}^u\right\|^2}
  {\left\|\lamsc \vec{a}^c+\lamsu\vec{a}^u\right\|^2}+
  \mathcal{O}(\axib^3)\;.
\end{equation}
We divide the index $\nu$ into two groups: The group $S$ in which
$\axinu$ are small and the expansions~\eqref{eq:43} and~\eqref{eq:44}
are justified, and the group $\overline{S}$ in which $\axinu \geq 1$
and the expansions are not justified. We are interested in the sum
over $\overline{S}$ only. We have
\begin{equation}\label{eq:49}
  \frac{\sum\limits_{\nu\in\overline{S}}\left|\lamsu\right|^2\left|a^u_{\nu}\right|^2}
  {\left\|\lamsc \vec{a}^c+\lamsu\vec{a}^u\right\|^2}
  \leq
  \frac{\sum\limits_{\nu\in\overline{S}}\left|\lamsu\right|^2\left|a^u_{\nu}\right|^2
    + \sum\limits_{\nu\in S}\left|\lamsu\right|^2\left|a^u_{\nu}\right|^2}
  {\left\|\lamsc \vec{a}^c+\lamsu\vec{a}^u\right\|^2} 
  = \axib^2 + \mathcal{O}(\axib^3)\;.
\end{equation}
However, in the group $\overline{S}$, where
$\left| \lamsc\, a^c_{\nu}\right| \leq
\left|\lamsu\,a^u_{\nu}\right|$, the full amplitude
$\left|A_{\nu}\right|^2$ can be at most
$4\left|\lamsu\,a^u_{\nu}\right|^2$. We therefore find that
\begin{equation}
  \label{eq:50}
  \frac{\sum\limits_{\nu\in\overline{S}}\left|A_{\nu}\right|^2}
  {\sum\limits_{\nu}\left|A_{\nu}\right|^2}
  \leq \frac{4\sum\limits_{\nu\in\overline{S}}\left|\lamsu\right|^2\left|a^u_{\nu}\right|^2}
  {\left\|\lamsc \vec{a}^c+\lamsu\vec{a}^u\right\|^2}
  \leq 4\axib^2 + \mathcal{O}(\axib^3)\;.
\end{equation}
One can easily see that omitting these small
$\left|A_{\nu\in\overline{S}}\right|^2$ terms from the sums in 
the expressions~\eqref{eq:39} and~\eqref{eq:40} corresponds to
omitting terms of ${\cal O}(\axib^2)$ in the expansions~\eqref{eq:43}
and~\eqref{eq:44}.

\section{The $\su$ Analysis}
\label{sec:su-ampl-relat}
Finding $SU(3)$ amplitude relations can be done systematically using
tensor methods. We write down the $(\pi,K,\eta)$ meson octet as
\begin{equation}\label{eq:51}
(P_8)^i_j = \begin{pmatrix}
\frac{1}{\sqrt{2}}\, \pi^0 +\frac{1}{\sqrt{6}}\, \eta_8 & \pi^+ & K^+ \\
\pi^- & -\frac{1}{\sqrt{2}}\, \pi^0 + \frac{1}{\sqrt{6}}\, \eta_8 & K^0 \\
 K^- & \overline{K}{}^0 & -\sqrt{\frac{2}{3}}\, \eta_8
  \end{pmatrix} \;,
\end{equation}
and the $B$ meson triplet as
\begin{equation}\label{eq:52}
(B_3)_i =\begin{pmatrix} B^+ & B_d & B_s \end{pmatrix}\;.
\end{equation}
We combine the $\Delta s=0$ and $\Delta s=-1$ Hamiltonian
operators into three rank $3$ tensors
\cite{Savage:1989ub,Dighe:1995bm}: 
\begin{align}\label{eq:53}
    ((H_3)^i)^j_k &= \left(
    \begin{pmatrix}
      0 & 0 & 0 \\
      0 & 0 & 0 \\
      0 & 0 & 0
    \end{pmatrix}\;,\right.
  \begin{pmatrix}
    \lambda^d_q & 0 & 0 \\
    0 & \lambda^d_q & 0 \\
    0 & 0 & \lambda^d_q
  \end{pmatrix}\;, \left.\begin{pmatrix}
      \lambda^s_q & 0 & 0 \\
      0 & \lambda^s_q & 0 \\
      0 & 0 & \lambda^s_q
    \end{pmatrix}
  \right)\;, \\
  \label{eq:54}
    ((H_{\bar{6}})^i)^j_k &= \left(
    \begin{pmatrix}
      0 & 0 & 0 \\
      \lambda^d_q & 0 & 0 \\
      \lambda^s_q & 0 & 0
    \end{pmatrix}\;,\right.
  \begin{pmatrix}
    -\lambda^d_q & 0 & 0 \\
    0 & 0 & 0 \\
    0 & -\lambda^s_q & \lambda^d_q
  \end{pmatrix}\;, \left.\begin{pmatrix}
      -\lambda^s_q & 0 & 0 \\
      0 & \lambda^s_q & -\lambda^d_q \\
      0 & 0 & 0
    \end{pmatrix}
  \right)\;, \\
  \label{eq:55}
    ((H_{15})^i)^j_k &= \left(
    \begin{pmatrix}
      0 & 0 & 0 \\
      3\lambda^d_q & 0 & 0 \\
      3\lambda^s_q & 0 & 0
    \end{pmatrix}\;,\right.
  \begin{pmatrix}
    3 \lambda^d_q & 0 & 0 \\
    0 & -2 \lambda^d_q & 0 \\
    0 & -\lambda^s_q & -\lambda^d_q
  \end{pmatrix}\;, \left. \begin{pmatrix}
      3 \lambda^s_q & 0 & 0 \\
      0 & - \lambda^s_q & -\lambda^d_q \\
      0 & 0 & -2 \lambda^s_q
    \end{pmatrix}
  \right)\;,
\end{align}
where $\lambda^{q^\prime}_q=V_{qb}^*V_{qq^\prime}$.

Usually, one proceeds by combining all mesons into irreducible
representations of $\su$ and contracting the Hamiltonian
operators in all possible ways. This would require a large amount of
multiplications of irreducible representations. Instead, we obtain a
set of independent reduced matrix elements by summing systematically
over all possible permutations: 
\begin{equation}
  \label{eq:56}
    \begin{split}
    \sum\limits_{\text{Permutations }p}
    \Bigg(\sum\limits_{i_1,\,i_2,\,i_3,\,i_4,\,i_5=1}^3 \Big[ &A_\mathbf{3}^\mathbf{p}\,
    (P_8)^{i_1}_{i_{p1}} (P_8)^{i_2}_{i_{p2}} (P_8)^{i_3}_{i_{p3}}
    (B_3)_{i_{p4}} (H_3)^{i_4\,i_5}_{i_{p5}} \\
    + & A_\mathbf{\bar{6}}^\mathbf{p}\, (P_8)^{i_1}_{i_{p1}} (P_8)^{i_2}_{i_{p2}}
    (P_8)^{i_3}_{i_{p3}}
    (B_3)_{i_{p4}} (H_{\bar{6}})^{i_4\,i_5}_{i_{p5}} \\
    +& A_\mathbf{15}^\mathbf{p}\, (P_8)^{i_1}_{i_{p1}} (P_8)^{i_2}_{i_{p2}}
    (P_8)^{i_3}_{i_{p3}} (B_3)_{i_{p4}} (H_{15})^{i_4\,i_5}_{i_{p5}}
    \Big]\Bigg)\;.
  \end{split}
\end{equation}
We remind the reader that the order of the final states mesons is
important as it corresponds to different momentum configurations.

All together there are $120$ permutations and therefore $360$
parameters ($A_\mathbf{3}^\mathbf{p}, \,A_\mathbf{\bar{6}}^\mathbf{p},
\,A_\mathbf{15}^\mathbf{p}$). However, these $360$ free parameters
appear in only a small number of combinations which
correspond to independent reduced matrix elements. Automating the
calculation, it is straightforward to obtain the set of such independent
combinations. Using Young diagrams, we verify that we obtain the
correct number of independent reduced matrix elements. We get $40$
independent reduced matrix elements: $10$ for $H_3$ in the
Hamiltonian (in other words, there are $10$ reduced matrix elements
$A^\mathbf{r_i}_\mathbf{3}$ where $\mathbf{r_i}$ stands for the $10$
different representations in $\mathbf{8}\times \mathbf{8} \times
\mathbf{8}$ which have non-zero reduced matrix elements involving
$H_3$), $12$ for $H_{\bar{6}}$, and $18$ for $H_{15}$.
The predictive power of the $\su$ symmetry is manifest when one
realizes that there are $95$ $\Delta s=0$ decays of $B^+$ and
$B^0$ decays to which we can relate the 3 $\Delta s=-1$ modes of
interest ($B^0\to\K\K\Kb$, $\K\Kb\K$ and $\Kb\K\K$).

In this work, only totally symmetric final states play a role. This
situation can be used to simplify the analysis. We replace the
combination $(P_8)^{i_1}_{i_{p1}} (P_8)^{i_2}_{i_{p2}}
(P_8)^{i_3}_{i_{p3}}$ in eq.~\eqref{eq:56} with the symmetrized
combination: 
\begin{equation}
  \label{eq:57}
  \begin{split}
    & (P_8)^{i_1}_{i_{p1}} (P_8)^{i_2}_{i_{p2}} (P_8)^{i_3}_{i_{p3}} +
    (P_8)^{i_2}_{i_{p2}} (P_8)^{i_3}_{i_{p3}} (P_8)^{i_1}_{i_{p1}} \\ 
    + & (P_8)^{i_3}_{i_{p3}} (P_8)^{i_1}_{i_{p1}} (P_8)^{i_2}_{i_{p2}} +
    (P_8)^{i_2}_{i_{p2}} (P_8)^{i_1}_{i_{p1}} (P_8)^{i_3}_{i_{p3}} \\ 
    + & (P_8)^{i_1}_{i_{p1}} (P_8)^{i_3}_{i_{p3}} (P_8)^{i_2}_{i_{p2}} +
    (P_8)^{i_3}_{i_{p3}} (P_8)^{i_2}_{i_{p2}} (P_8)^{i_1}_{i_{p1}} \;.
  \end{split}
\end{equation}
Then the number of independent matrix elements is reduced to
$7$. The predictive power is maintained, since there are 20 
symmetrized $\Delta s=0$ modes to which we relate the single
$B^0\to\SK$ decay amplitude. Note, however, that the symmetric
states are not properly normalized and so their normalization needs to
be introduced by hand (see eq.~\eqref{eq:5}). 

Table~\ref{tab:matrix-elements} lists the independent reduced matrix
elements. The names of the matrix elements ($A_1$, $A_2$ etc.) bear no
significance. The horizontal lines divide the table according to the
form of the symmetrized final state. The symmetrized states in the
table are all normalized. 

In this work we also consider a simplifying dynamical assumption by
which we neglect the effect of exchange, annihilation and penguin
annihilation diagram~\cite[]{Dighe:1995bm}. The implementation of this
assumption is straightforward in our calculation. One should just drop all
permutations in which the $B$-meson triplet is contracted with the Hamiltonian
operator. In other words, one takes eq.~(\ref{eq:56}) and drops from
the sum all permutations in which $p4=4$ or $p4=5$. When this
procedure is applied to the symmetrized combination~(\ref{eq:57}),
there are only $5$ independent reduced matrix elements.

\section{Experimental data}
\label{sec:exp-data}

We quote experimental data relevant to three pseudoscalar final
states. Measurements where resonant contributions are removed from the
sample are denoted by (NR). The currently measured $\Delta s=\pm1$ modes
are~\cite{HFAG}: 
\begin{equation}
  \label{eq:58}
  \begin{aligned}
    \mathcal{B}({K_SK_SK_S}) &= (5.8 \pm 1.0) \times
    10^{-6},\\
    \mathcal{B}({K^+ \pi^+ \pi^-}) &= (53.5 \pm 3.5) \times
    10^{-6},\\
    \mathcal{B}^{(NR)}({K^+ \pi^+ \pi^-}) &= (4.9 \pm 1.5) \times
    10^{-6},\\
    \mathcal{B}({K^+ K^- K^+}) &= (30.1 \pm 1.9) \times
    10^{-6},\\
    \mathcal{B}({K^+ K_S K_S}) &= (11.5 \pm 1.3) \times
    10^{-6},\\
    \mathcal{B}({\eta K^+ \pi^-}) &= (33.4^{+4.1}_{-3.8}) \times
    10^{-6},\\
    \mathcal{B}({K^0 \pi^+ \pi^-}) &= (44.9 \pm 4.0) \times
    10^{-6},\\
    \mathcal{B}({K^+ \pi^- \pi^0}) &= (35.6^{+3.4}_{-3.3}) \times
    10^{-6},\\
    \mathcal{B}({K^+ K^- K^0}) &= (24.7 \pm 2.3) \times
    10^{-6}.
  \end{aligned}
\end{equation}
The currently measured or constrained $\Delta s=0$ modes
are~\cite{HFAG,Aubert:2005dy,Eidelman:2004wy}:

\begin{equation}
  \label{eq:59}
  \begin{aligned}
    \mathcal{B}({\pi^+ \pi^- \pi^+}) &= (16.2 \pm 2.5) \times
    10^{-6},\\
    \mathcal{B}({\pi^+ \pi^- \eta}) &= (16.6^{+3.8}_{-3.4}) \times
    10^{-6},\\
    \mathcal{B}({K^+ K^- \pi^+}) &< 6.3 \times
    10^{-6},\\
    \mathcal{B}({K_S K_S \pi^+}) &< 3.2 \times
    10^{-6},\\
    \mathcal{B}({K^+ \overline{K}{}^0 \pi^0}) &< 24 \times
    10^{-6},\\
    \mathcal{B}({\K K^- \pi^+}) &< 21.0 \times
    10^{-6}, \\
    \mathcal{B}({K^+ K^- \pi^0}) &< 19 \times
    10^{-6},\\
    \mathcal{B}({K^+ \overline{K}{}^0 \pi^-}) &< 18 \times
    10^{-6},\\
    \mathcal{B}^{(NR)}({\pi^+ \pi^- \pi^0}) &< 7.3 \times
    10^{-6}.
\end{aligned}
\end{equation}


\end{document}